# Laws of 4D printing

Farhang Momeni* and Jun Ni

Department of Mechanical Engineering, University of Michigan, Ann Arbor, MI 48109, USA.

*Corresponding author; e-mail: farhang@umich.edu

**Abstract**

The main difference between 3D and 4D printed structures is one extra dimension that is smart evolution over "time". However, currently, there is no general formula to model and predict this extra dimension. Here, by starting from fundamental concepts, we derive and validate a "universal" bi-exponential formula that "is required" to model and predict the fourth D of 4D printed multi-material structures. 4D printing is a new manufacturing paradigm to elaborate stimuli-responsive materials in multi-material structures for advanced manufacturing (and construction) of advanced products (and structures). It conserves the general attributes of 3D printing (such as the elimination of molds, dies, and machining) and further enables the fourth dimension of products and structures to provide intelligent behavior over time. This intelligent behavior is encoded (usually by an inverse mathematical problem) into stimuli-responsive multi-materials during printing and is enabled by stimuli after printing. Here, we delve into the fourth dimension and reveal three general laws that govern the time-dependent shape-shifting behaviors of almost "all" (photochemical-, photothermal-, solvent-, pH-, moisture-, electrochemical-, electrothermal-, ultrasound-, enzyme-, etc.-responsive) "multi-material" 4D structures. We demonstrate that two different types of time-constants govern the shape-shifting behavior of almost all the multi-material 4D printed structures over time. Our results starting from the most fundamental concepts and ending with governing equations can serve as general design principles for future research in the 4D printing field, where the "time-dependent" behaviors should be understood, modeled, and predicted, correctly. Future software and hardware developments in 4D printing can also benefit from these results.

## 1. Introduction

4D printing [1-6] is the art of combining smart (active, stimuli-responsive) materials [7-9], mathematics, and multi-material additive manufacturing. These three elements yield an encoded, multi-material (multi-component), single-piece, smart structure that demonstrates predictable and desired dynamic intelligent behaviors under the right stimulus through an interaction mechanism [6].

Although some studies demonstrated single-material 4D printed structures, the future of 4D printing will lie in multi-material structures [2, 10-17]. The quiddity of 4D printing usually needs multi-material structures [11]. In the most fundamental case, the multi-material 4D printed structure has one active and one passive layer [12]. Performance improvement by allocating proper materials to related locations based on local necessities [18], multi-functionality by embeddable functions such as electronics [19], combing rigid and flexible sections in an integrated structure [20], and providing lightweight structures [21] are only some of the advantages of multi-material structures. In addition, to enable the shape memory effect (SME) (that is *not* an intrinsic property [9]) of shape memory polymers (SMPs), a mechanical force is required in addition to heat (thermomechanical cycles). This force is usually provided externally. However, 4D printing can help us to arrange active and passive materials in a multi-material structure and utilize their internal mismatch-driven forces to enable the SME autonomously, with no need of external load for training [22, 23]. In this study, we focus on "multi-material" 4D printed structures.

Aspects of 4D printing have been explored in the literature. Several studies worked on beam and plate theories (e.g., refer to the supplementary information of ref. [5]). However, the missing piece in the literature is modeling of "time-dependent" behaviors (the $4^{th}$ D) of 4D structures. Especially, the time-dependent behavior is the critical part of 4D (stimuli-responsive) materials, whether fabricated by additive manufacturing and thus called 4D *printed* structures or created by other manufacturing processes. More importantly, a huge number of studies on 4D materials used the Timoshenko bimetal model [24] (that is linear with time) to analyze the time-*dependent* behaviors of their experiments. Here, we will see that, in general, the Timoshenko bimetal model cannot capture the true time-*dependent* behaviors of 4D materials (except for some special cases or selected linear regions), although it provides useful insights on time-*independent* behaviors. In fact, the purpose of Timoshenko bimetal model was not to model the time-*dependent* behavior (the $4^{th}$ D) of 4D materials. Thus, there is an urgent need for qualitative and quantitative analysis on the fourth D of 4D materials.

The main part of 4D printed structures is the $4^{th}$ D; however, currently, there is no general formula to model and predict this extra dimension. Here, by developing fundamental concepts and from the equilibrium and compatibility conditions, we derive a bi-exponential formula that "is needed" for modeling and predicting the $4^{th}$ D of any multi-material 4D printed structure. We further validate our bi-exponential formula by various experimental data from separate studies in the literature and show that it is a general formula that is useful for any type of 4D multi-material structure (photochemical-, photothermal-, solvent-, pH-, moisture-, electrochemical-, electrothermal-, ultrasound-, enzyme-, etc.-responsive). This generality happens, because we build the bases of our bi-exponential formula, comprehensively.

2. **Results and discussions**

There are many materials and stimuli. Consequently, most of the ongoing studies in the 4D printing field are case-specific. The time-dependent behavior (the fourth D) of any 4D printed structure needs to be predicted. A detailed, but systematic, study on 4D printing and related areas helped us to reveal three universal laws that govern the "shape-shifting" behaviors of almost all the "multi-material" 4D printed structures, although there are many materials and stimuli (Figure 1).

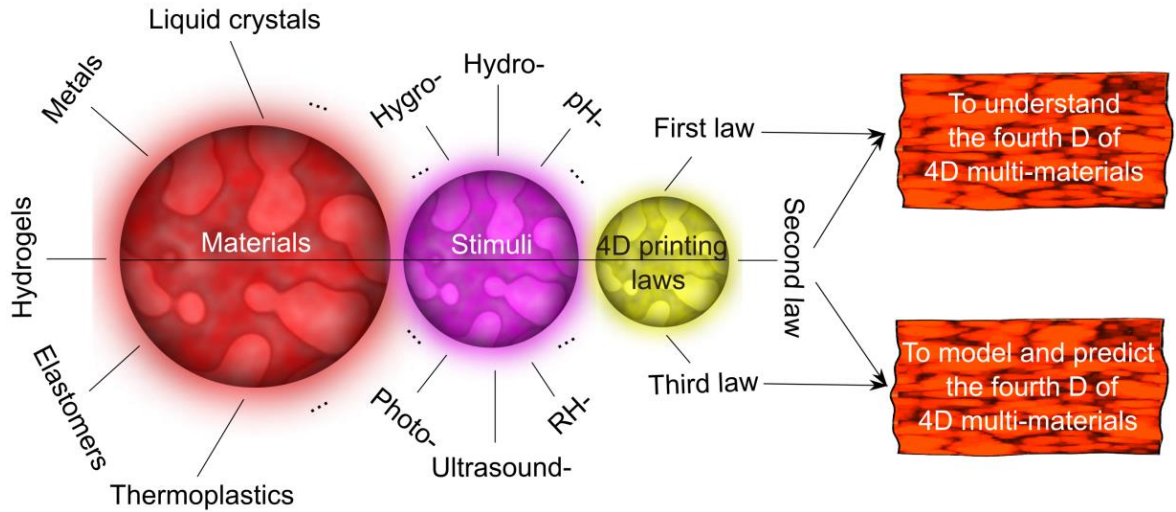

Figure 1. Toward the 4D printing laws.

## 2.1. First law of 4D printing

*Almost all the shape-shifting behaviors (photochemical-, photothermal-, solvent-, pH-, moisture-, electrochemical-, electrothermal-, ultrasound-, enzyme-, hydro-, thermo-, etc.-responsive) of the "multi-material" 4D printed structures originate from one fundamental phenomenon that is "relative expansion" between active and passive materials.*

This "relative expansion" is the origin of almost all the complicated 4D printing shape-shifting behaviors such as twisting, coiling, curling, etc., that are enabled by encoding various types of anisotropy between active and passive materials and fabricating different heterogeneous structures.

## 2.2. Second law of 4D printing

*The shape-shifting behaviors of almost all the "multi-material" 4D printed structures have four different types of physics; mass diffusion, thermal expansion/contraction, molecular transformation, and organic growth. They all (discussed and quantified below) lead to the relative expansion between active and passive materials and consequent shape-shifting behaviors under stimuli (the stimulus is usually provided externally, but it can be internal).*

### 2.2.1. Quantifying the second law

Here, we describe and quantify the four underlying physical concepts that lead to relative expansions between active and passive materials in "multi-material" 4D structures, resulting in various shape-shifting behaviors, with or without shape memory effect.

#### 2.2.1.1. Mass Diffusion

In this category, a matter transport leads to the relative expansion.

Mass change due to sorption (absorption or adsorption) of a guest matter (here is mainly a stimulus such as water, ion, etc.) in a host matter can be modeled as below [25-27]:

$$\frac{\Delta M}{M}(t) = C\left[1 - \exp\left(-\frac{t}{\tau}\right)\right],$$

where $t$ is time and $M$ is mass. $C$ and $\tau$ are usually obtained by curve fitting to experimental data and depend on host matter relaxation and guest matter diffusion. However, models can be developed for these two parameters. The exponential model above captures the correct behavior of mass diffusion for short- to long-time processes. There is also one other model called power function ($kt^n$), which is not accurate for long-time processes of mass diffusion [25]. However, we know that the main part of 4D printing is the intelligent behavior over "time" that can be short or long.

The exponential model above has been mainly introduced for $\frac{\Delta M}{M}$ and $\frac{\Delta V}{V}$ (volumetric strain). However, we have

$$\frac{\Delta V}{V} = \frac{V_2 - V_1}{V} = \frac{(L_x + \Delta L_x)(L_y + \Delta L_y)(L_z + \Delta L_z) - L_x L_y L_z}{L_x L_y L_z} \cong \varepsilon_x + \varepsilon_y + \varepsilon_z,$$

where $V$ is volume, $L$ is length, $\varepsilon$ is strain, and the second- and third-order differential quantities are neglected.
For an isotropic material,

$$\varepsilon_x = \varepsilon_y = \varepsilon_z = \varepsilon \Rightarrow \varepsilon = \frac{1}{3}\frac{\Delta V}{V}.$$

In addition, mass and volume have a linear relationship. Therefore,

$$\varepsilon_{\text{Mass Diffusion}}(t) = C_1\left[1 - \exp\left(-\frac{t}{\tau_1}\right)\right],$$

where $C_1$ and $\tau_1$ are constants that depend on the previous parameters.

We quantify all the four categories in terms of strain. One reason is that for example, if we want to quantify them in terms of mass, then we do not have any mass change in the next category (Thermal Expansion/Contraction). Similarly, we do not have temperature change in this category, while we have it in the next category.

In this category, several stimuli can be used such as hydro-, solvent-, moisture-, pH-, enzyme-, photochemical-, and electrochemical-responsive mechanisms. All of these stimuli finally cause a matter transport, leading to relative expansion in multi-material 4D structures, resulting in various shape-shifting behaviors.

### 2.2.1.2. Thermal Expansion/Contraction

In this category, a temperature change will increase (or decrease) the average distance between atoms and molecules (with constant mass), leading to the relative expansion in multi-materials.

Strain due to temperature change is [28]:

$$\varepsilon_{\text{thermal}} = \alpha \Delta T,$$

where $\alpha$ is thermal expansion coefficient and $T$ is temperature. Because we need to predict the behavior of 4D printed structures over *time*, we convert this temperature-based equation to time-based form.

On the other hand, by applying a thermal stimulus with temperature $T_2$ to a structure with temperature $T_1$, and assuming a uniform temperature in the structure, the temperature of the structure changes over time as below [29]:

$$T_1(t) = T_2 + [T_1(t=0) - T_2]\exp\left(-\frac{t}{\tau}\right) + R(\dot{S}_1 - Q_1)\left[1 - \exp\left(-\frac{t}{\tau}\right)\right],$$

where $t$ is time, $T_1(t=0)$ is the initial temperature of the structure, $\dot{S}_1$ is energy conversion in the structure, $Q_1$ is the heat transferred between the structure and environment (other than the stimulus), and $R$ is thermal resistance. $\tau$ is a time-constant that depends on density, heat capacity, volume of the structure, and thermal (and thermal contact) resistance, and can be modeled in a specific application [29].

By working on the above equation, we have

$$T_1(t) - T_1(t=0) = [T_2 - T_1(t=0)] - [T_2 - T_1(t=0)]\exp\left(-\frac{t}{\tau}\right) + R(\dot{S}_1 - Q_1)\left[1 - \exp\left(-\frac{t}{\tau}\right)\right],$$

$$\Rightarrow T_1(t) - T_1(t=0) = [T_2 - T_1(t=0)]\left[1 - \exp\left(-\frac{t}{\tau}\right)\right] + R(\dot{S}_1 - Q_1)\left[1 - \exp\left(-\frac{t}{\tau}\right)\right],$$

$$\Rightarrow T_1(t) - T_1(t=0) = \left[T_2 - T_1(t=0) + R(\dot{S}_1 - Q_1)\right]\left[1 - \exp\left(-\frac{t}{\tau}\right)\right],$$

$$\Rightarrow \Delta T_1 = \left[T_2 - T_1(t=0) + R(\dot{S}_1 - Q_1)\right]\left[1 - \exp\left(-\frac{t}{\tau}\right)\right].$$

By combining the above equation and the initial equation (i.e., $\varepsilon = \alpha \Delta T$), we will have

$$\varepsilon_{\text{Thermal Expansion/Contraction}}(t) = C_2\left[1 - \exp\left(-\frac{t}{\tau_2}\right)\right],$$

where $C_2$ and $\tau_2$ are constants that depend on the previous parameters.

In this category, several stimuli can be used such as photothermal-, electrothermal-, and ultrasound-responsive mechanisms. All of these stimuli finally raise the temperature of the structure and consequently, increase the average distances between atoms and molecules. For example, in electrothermal-responsive structures, the movement of a current through a resistance provides heat (so-called Joule or Ohmic heating), this heat increases the temperature, and finally, the expansion happens. Similarly, contraction is obtained by cooling. It should be noted that some materials will shrink (contract) upon heating. In these cases, $\alpha$ (the thermal expansion coefficient) would be negative in the equations above and the final equation, i.e., $\varepsilon_{\text{Thermal Expansion/Contraction}}(t) = C_2\left[1 - \exp\left(-\frac{t}{\tau_2}\right)\right]$ will remain intact. Nevertheless, the key point is the "relative" thermal expansion/shrinkage in multi-materials to enable various shape-shifting behaviors in this category.

### 2.2.1.3. Molecular Transformation

In this category, a molecular transformation (e.g., trans-to-cis change in azobenzene) with constant mass and temperature leads to the relative expansion. The constant-temperature feature of this category indicates that the relative expansion is not due to the thermal expansion/contraction, even though some thermal fluctuation may happen in the structure due to bond cleavage or formation.

The molecular transformation is usually accomplished by photochemical responsivity mechanism that is different from photothermal responsivity [30]. The kinetics of photoinduced transformation is [30]:

$$D(t) = D_0 \left[1 - \exp\left(-\frac{t}{\tau}\right)\right],$$

where $t$ is time and $D(t)$ is the degree of transformation. $D_0$ and $\tau$ are constants that depend on the irradiation flux intensity, quantum yield of the transformation, etc., and can be modeled in a specific application [30].

On the other hand, subsequent volume change (expansion or contraction) is proportional to the degree of transformation [30]. In addition, we have already seen that the volumetric strain ($\frac{\Delta V}{V}$) and linear strain ($\varepsilon = \frac{\Delta L}{L}$) are proportional. Therefore,

$$\varepsilon_{\text{Molecular Transformation}}(t) = C_3 \left[1 - \exp\left(-\frac{t}{\tau_3}\right)\right],$$

where $C_3$ and $\tau_3$ are constants that depend on the previous parameters.

The main stimulus for this category is the photochemical-responsive mechanism.

### 2.2.1.4. Organic Growth

In this category, there is a living layer (organism) and its growth over time can lead to the relative expansion between active and passive materials. The organic growth can be defined as the increase of an organic system in weight or length [31]. This usually happens in bioscience and bioengineering dealing with cells, soft tissues, organs, scaffolds, and so on that can be 4D printed and are generally called 4D bioprinting.

Kinetics of the organic growth is [31]:

$$L(t) = L_\infty - (L_\infty - L_0) \exp\left(-\frac{t}{\tau}\right),$$

where $t$ is time, $L(t)$ is the length of the organic system, $L_\infty$ is the final length, and $L_0$ is the initial length. $\tau$ is usually a curve-fitting time-constant that depends on the metabolism of the living organism, environment, and so on. Nevertheless, models can be developed for it. It should also be noted that $\tau$ affects $L_\infty$ [31]. This formula shows the growth of *individual* organisms in a population that is different from *population* growth. The population growth is the growth in the number of individuals in a population and is modeled by other formulas [32, 33].

On the other hand, based on the definition of strain, $\varepsilon = \frac{L(t) - L_0}{L_0}$, we have

$$\varepsilon = \frac{L(t) - L_0}{L_0} = \frac{(L_\infty - L_0) - (L_\infty - L_0) \exp\left(-\frac{t}{\tau}\right)}{L_0} = \left(\frac{L_\infty - L_0}{L_0}\right)\left[1 - \exp\left(-\frac{t}{\tau}\right)\right].$$

Therefore,

$$\varepsilon_{\text{Organic Growth}}(t) = C_4 \left[1 - \exp\left(-\frac{t}{\tau_4}\right)\right],$$

where $C_4$ and $\tau_4$ are constants that depend on the previous parameters.

In this category, one of the main stimuli that can trigger a living organism would be the electrical signal (electrochemical mechanism). In addition, various stimuli such as pH, light, heat, and enzyme can be used to tune the growth rate.

### 2.2.2. Unified model of the second law

By quantifying the second law, we found that

$$\begin{cases} \varepsilon_{\text{Mass Diffusion}}(t) = C_1\left(1 - e^{-\frac{t}{\tau_1}}\right) \\ \varepsilon_{\text{Thermal Expansion/Contraction}}(t) = C_2\left(1 - e^{-\frac{t}{\tau_2}}\right) \\ \varepsilon_{\text{Molecular Transformation}}(t) = C_3\left(1 - e^{-\frac{t}{\tau_3}}\right) \\ \varepsilon_{\text{Organic Growth}}(t) = C_4\left(1 - e^{-\frac{t}{\tau_4}}\right) \end{cases},$$

where $C_i$ and $\tau_i$ ($i = 1,2,3,4$) all are constants. However, they depend on different factors as described.

## 2.3. Third law of 4D printing

*Time-dependent shape-shifting behavior of almost all the "multi-material" 4D printed structures is governed by two "different types" of time-constants. For the most fundamental case of a multi-material 4D printed structure having one active and one passive layer (Figure 2), the time-dependent behavior (in terms of "curvature" that is a building block concept for shape-shifting) is*

$$\kappa(t) = \frac{1}{\rho}(t) = \frac{H_I\left(1 - e^{\frac{-t}{\tau_I}}\right) + H_{II}\left(1 - e^{\frac{-t}{\tau_{II}}}\right)}{\frac{h}{2} + \frac{2(E_1 I_1 + E_2 I_2)}{h}\left(\frac{1}{E_1 a_1} + \frac{1}{E_2 a_2}\right)} = K_I\left(1 - e^{\frac{-t}{\tau_I}}\right) + K_{II}\left(1 - e^{\frac{-t}{\tau_{II}}}\right), \quad (1)$$

where $t$ is time, $\kappa(t)$ is the curvature induced by the relative expansion, $\rho$ is the radius of curvature, $h$ and $a_i$ are thicknesses identified in Figure 2, $E_i$ is Young`s modulus, and $I_i$ is the second moment of area. The passive and active layers are denoted by numbers 1 and 2, respectively. $H_I$ is a constant that depends on Young`s moduli of the active and passive layers and the amount of the mismatch-driven stress generated at the interface. $\tau_I$ is a time-constant that depends on the viscosity induced at the interface and Young`s moduli of the active and passive layers. The viscosity induced at the interface needs to be measured or modeled for a specific active-passive composite. The exact format of $H_I$ and $\tau_I$ can be developed in a specific application depending on the active-passive composite and by using parallel and series rules of springs (elasticity elements) and dashpots (viscosity elements). $H_{II}$ and $\tau_{II}$ are respectively equivalent to $C_i$ and $\tau_i$ ($i = 1,2,3,4$) of the unified model in the previous section.

### 2.3.1. Proof of the third law

To derive equation (1), we start from the equilibrium and compatibility conditions that are the starting point for any problem in mechanics of materials [28]. We also consider the Timoshenko bimetal model [24] (and its basic assumptions).

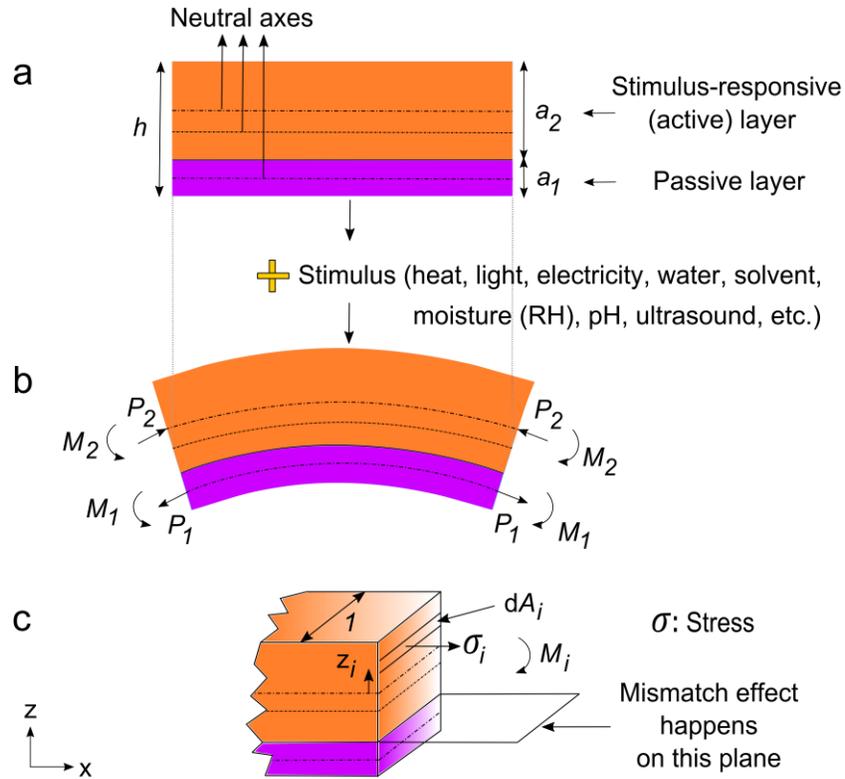

Figure 2. Toward the third law by analyzing the most fundamental multi-material 4D structure.

**Equilibrium:**

First, we must have balances of forces and moments in Figure 2. Therefore [24],

$$Balance\ of\ forces: \sum F = 0 \Rightarrow P_1 = P_2 = P \qquad (2)$$

$$Balance\ of\ moments: \sum M = 0 \Rightarrow \frac{P_1 a_1}{2} + \frac{P_2 a_2}{2} = M_1 + M_2 \xRightarrow[a_1+a_2=h]{P_1=P_2=P} \frac{Ph}{2} = M_1 + M_2, \qquad (3)$$

where $P_i$ and $M_i$ are forces and moments, respectively (shown in Figure 2).

**Compatibility:**

Second, *at the interface* of the two layers, the lengths of the two layers are the same after applying the stimulus. Because their initial lengths are also the same, their strains must be equal and thus [24, 28],

$$At\ the\ interface: \varepsilon_1 = \varepsilon_2.$$

The strain in each of the two layers has three main contributors as below [24, 34].

$$\left(\varepsilon_{curvature} + \varepsilon_{stress} + \varepsilon_{expansion}\right)_1 = \left(\varepsilon_{curvature} + \varepsilon_{stress} + \varepsilon_{expansion}\right)_2. \qquad (4)$$

**Strain from curvature:** $\varepsilon_{curvature}$

$$\begin{cases} (\varepsilon_{curvature})_1 = \dfrac{a_1}{2\rho} \\ (\varepsilon_{curvature})_2 = -\dfrac{a_2}{2\rho} \end{cases} [24]. \qquad (5)$$

**Strain from stress:** $\varepsilon_{stress}$

$$\begin{cases} (\varepsilon_{stress})_1 = \dfrac{P_1}{E_1 a_1} \\ (\varepsilon_{stress})_2 = -\dfrac{P_2}{E_2 a_2} \end{cases} [24]. \qquad (6)$$

$$\xrightarrow{(4),(5),\text{and }(6)} \quad \frac{a_1}{2\rho} + \frac{P_1}{E_1 a_1} + (\varepsilon_{expansion})_1 = -\frac{a_2}{2\rho} - \frac{P_2}{E_2 a_2} + (\varepsilon_{expansion})_2. \qquad (7)$$

Now, in the following, we develop $\varepsilon_{stress}$ and $\varepsilon_{expansion}$ for 4D multi-materials and incorporate them in the equilibrium and compatibility equations.

Here, we note that $\varepsilon_{stress}$ is the strain due to the *mismatch-driven stress* at the interface of the active and passive materials. The mismatch-driven stress naturally leads to opposing resistive forces in the two layers (in general). We include the resistive effect by expanding the well-known moment equation for each layer, $M_i$. By considering Figure 2(c),

$$M_i = \int \boldsymbol{r_i} \times d\boldsymbol{F_i} = \int \boldsymbol{z_i} \sigma_i dA_i = \left(\int \boldsymbol{z_i} \sigma_i dA_i\right)_{except\ interface} + \left(\int \boldsymbol{z_i} \sigma_i dA_i\right)_{interface}. \qquad (8)$$

Because the integral is the same as the summation, i.e., it is continuous summation, we could separate the integral above into two terms as equation (8). The second term on the right-hand side of (8) shows the integral over an *infinitesimal* cross-sectional area, $dA_i$ (Figure 2(c)) that is close to the interface. Therefore, equation (8) can be written as

$$\begin{cases} M_1 = \dfrac{E_1 I_1}{\rho} + m_1 \\ M_2 = \dfrac{E_2 I_2}{\rho} + m_2 \end{cases}. \qquad (9)$$

The first terms on the right-hand side of (9) are similar to those proposed by Timoshenko [24] and the second terms ($m_1$ and $m_2$) arise from the resistive (mismatch) effect at the interface of the active and passive materials. At this stage, the nature of $m_1$ and $m_2$ is moment. Let us keep them as black-box terms.

**Strain from expansion that is enabled by stimulus:** $\varepsilon_{expansion}$

By analyzing almost all types of shape-shifting mechanisms in multi-material 4D structures, the relative expansions induced under stimuli can be categorized into four main groups elaborated in the second law. We demonstrated that almost all types of strains due to expansions induced by stimuli have the same format as below:

$$\varepsilon_{expansion}(t) = C\left(1 - e^{\frac{-t}{\tau_{II}}}\right). \tag{10}$$

We use $\tau_{II}$ as the time-constant of $\varepsilon_{expansion}$ to distinguish it from $\tau_I$ that will be introduced for the strain due to the *mismatch-driven* stress ($\varepsilon_{stress}$).

By combining equations (2), (3), and (7),

$$\frac{h}{2\rho} + \frac{2(M_1 + M_2)}{h}\left(\frac{1}{E_1 a_1} + \frac{1}{E_2 a_2}\right) = (\varepsilon_{expansion})_2 - (\varepsilon_{expansion})_1 \stackrel{(9)}{\Rightarrow}$$

$$\frac{h}{2\rho} + \frac{2(E_1 I_1 + E_2 I_2)}{h\rho}\left(\frac{1}{E_1 a_1} + \frac{1}{E_2 a_2}\right) + Nm_1 + Nm_2 = \underbrace{(\varepsilon_{expansion})_2 - (\varepsilon_{expansion})_1}_{\varepsilon_{relative\ expansion}},$$

where $N = \frac{2}{h}\left(\frac{1}{E_1 a_1} + \frac{1}{E_2 a_2}\right)$. Now, we assume that the active (stimuli-responsive) material is usually responsive under stimulus, and the passive material is not usually responsive under stimulus, as their names imply (nevertheless, the relative expansion is important for shape-shifting). Therefore, by applying equation (10) to the above equation,

$$\frac{h}{2\rho} + \frac{2(E_1 I_1 + E_2 I_2)}{h\rho}\left(\frac{1}{E_1 a_1} + \frac{1}{E_2 a_2}\right) + Nm_1 + Nm_2 = C\left(1 - e^{\frac{-t}{\tau_{II}}}\right).$$

Now, each term in the above equation is strain. Therefore, the nature of $Nm_1$ and $Nm_2$ is strain. On the other hand, these two terms reflect the mismatch (viscosity) effect of the interface into each layer. The viscoelastic *strain* over *time* can be modeled by an exponential term as below [35].

$$\frac{h}{2\rho} + \frac{2(E_1 I_1 + E_2 I_2)}{h\rho}\left(\frac{1}{E_1 a_1} + \frac{1}{E_2 a_2}\right) + A_1\left(1 - e^{\frac{-t}{B_1}}\right) + A_2\left(1 - e^{\frac{-t}{B_2}}\right) = C\left(1 - e^{\frac{-t}{\tau_{II}}}\right),$$

where $A_1$ and $A_2$ are constants that depend on Young`s moduli of the active and passive layers and the amount of the mismatch-driven stress generated at the interface. This stress is affected by the stimulus power (such as light intensity, pH value, etc.). $B_1$ and $B_2$ are constants that depend on the viscosity induced at the interface (that is related to the active-passive composite) and Young`s moduli of the active and passive layers. It should be noted that the Young`s modulus and viscosity are affected by the fabrication process [23] and its conditions (such as printing resolution).

On the other hand, because the two layers are attached at the interface during the shape-shifting, $B_1$ and $B_2$ (time-constants of strains in each layer due to the mismatch-driven stress at the interface) are equal ($B_1 = B_2 = \tau_I$). Thus,

$$\frac{h}{2\rho} + \frac{2(E_1 I_1 + E_2 I_2)}{h\rho}\left(\frac{1}{E_1 a_1} + \frac{1}{E_2 a_2}\right) + A_1\left(1 - e^{\frac{-t}{\tau_I}}\right) + A_2\left(1 - e^{\frac{-t}{\tau_I}}\right) = C\left(1 - e^{\frac{-t}{\tau_{II}}}\right).$$

By using new uniform notations $H_I$ and $H_{II}$,

$$\frac{h}{2\rho} + \frac{2(E_1I_1 + E_2I_2)}{h\rho}\left(\frac{1}{E_1a_1} + \frac{1}{E_2a_2}\right) = H_I\left(1 - e^{\frac{-t}{\tau_I}}\right) + H_{II}\left(1 - e^{\frac{-t}{\tau_{II}}}\right).$$

Finally, by re-arranging,

$$\frac{1}{\rho} = \frac{H_I\left(1 - e^{\frac{-t}{\tau_I}}\right) + H_{II}\left(1 - e^{\frac{-t}{\tau_{II}}}\right)}{\frac{h}{2} + \frac{2(E_1I_1 + E_2I_2)}{h}\left(\frac{1}{E_1a_1} + \frac{1}{E_2a_2}\right)} = K_I\left(1 - e^{\frac{-t}{\tau_I}}\right) + K_{II}\left(1 - e^{\frac{-t}{\tau_{II}}}\right),$$

which is the same as equation (1).

### 2.3.2. Stimulus-on versus stimulus-off

Equation (1) is used when the stimulus is "on", and a curvature happens in the structure. However, when the stimulus is "off", the structure can return to its original shape by starting from the final curvature of the previous part (i.e., the stimulus-on region). Therefore, the governing equation for the second region can be found as below:

$$\left(\frac{1}{\rho}\right)_{off} = \left\{\lim_{\substack{t \to \infty \\ (\text{or large } t)}}\left[K_I\left(1 - e^{\frac{-t}{\tau_I}}\right) + K_{II}\left(1 - e^{\frac{-t}{\tau_{II}}}\right)\right]\right\} - \left[K_I\left(1 - e^{\frac{-t}{\tau_I}}\right) + K_{II}\left(1 - e^{\frac{-t}{\tau_{II}}}\right)\right]$$

$$= (K_I + K_{II}) - \left[K_I\left(1 - e^{\frac{-t}{\tau_I}}\right) + K_{II}\left(1 - e^{\frac{-t}{\tau_{II}}}\right)\right]$$

$$= K_I e^{\frac{-t}{\tau_I}} + K_{II} e^{\frac{-t}{\tau_{II}}}. \tag{11}$$

It should be noted that in some applications, a self-locking mechanism could be devised by special arrangements of active and passive materials so that when the stimulus is off, the structure does not return to its original shape.

### 2.3.3. General graph

Based on equations (1) and (11), the general graph is rendered in Figure 3.

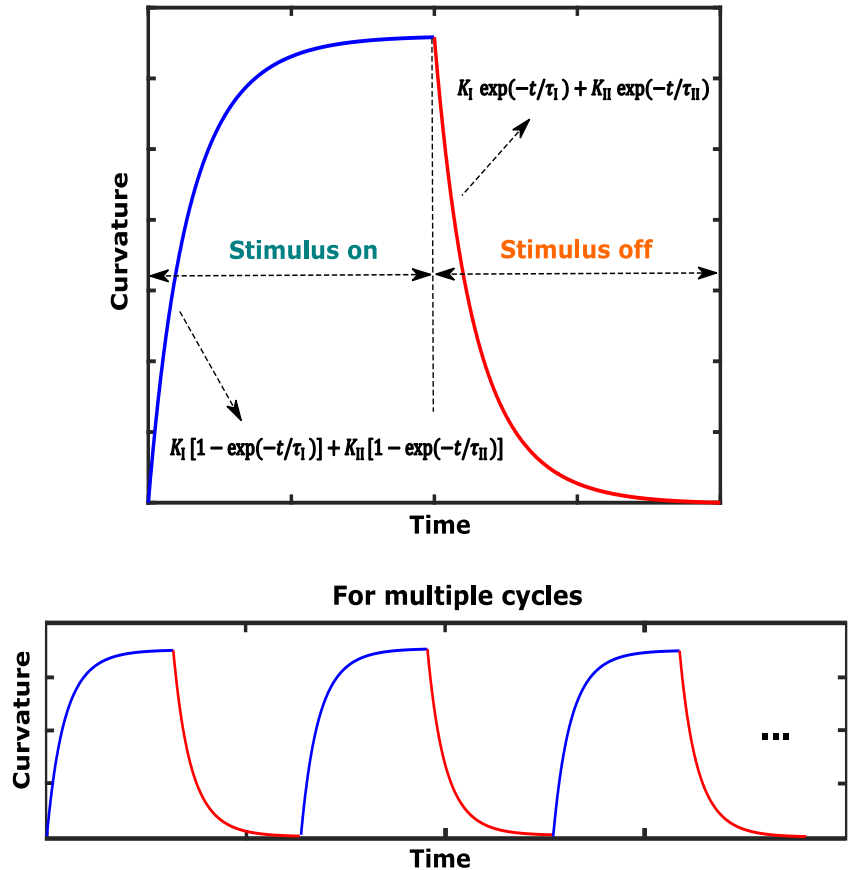

Figure 3. The general graph that exhibits the time-dependent behavior of almost all the multi-material 4D printed structures (photochemical-, photothermal-, solvent-, pH-, moisture-, electrochemical-, electrothermal-, ultrasound-, enzyme-, hydro-, etc.-responsive). Some applications need only one cycle, and some others require multiple cycles. In some applications, only one of the two regions of the graph happens and in some other applications, both the regions are present. In some cases, the shape-shifting behavior can occur with memory (SME), and in some other cases, it can take place without memory.

### 2.3.4. Validation

Here, we validate our derived bi-exponential formula. For completeness, we also consider the Timoshenko bimetal [24] and mono-exponential models. As seen in Figure 4, unlike the Timoshenko bimetal and mono-exponential models, the developed bi-exponential model perfectly captures the correct time-dependent behavior of various experimental data from separate studies in the literature. Therefore, in general, the time-dependent behavior of 4D multi-materials is nonlinear (with time) and has the specific format as equation (1). Some of the following studies presented experimental data for time-dependent curvature, and some others provided experimental data for time-dependent angle of rotation (deflection angle). However, the curvature and deflection angle have a linear relationship. Consequently, if one of them has bi-exponential behavior, the other one will have bi-exponential behavior. All the six items in Figure 4 have one active material and one passive material. The active component has responsivity to the desired stimulus, whether with or without shape memory effect.

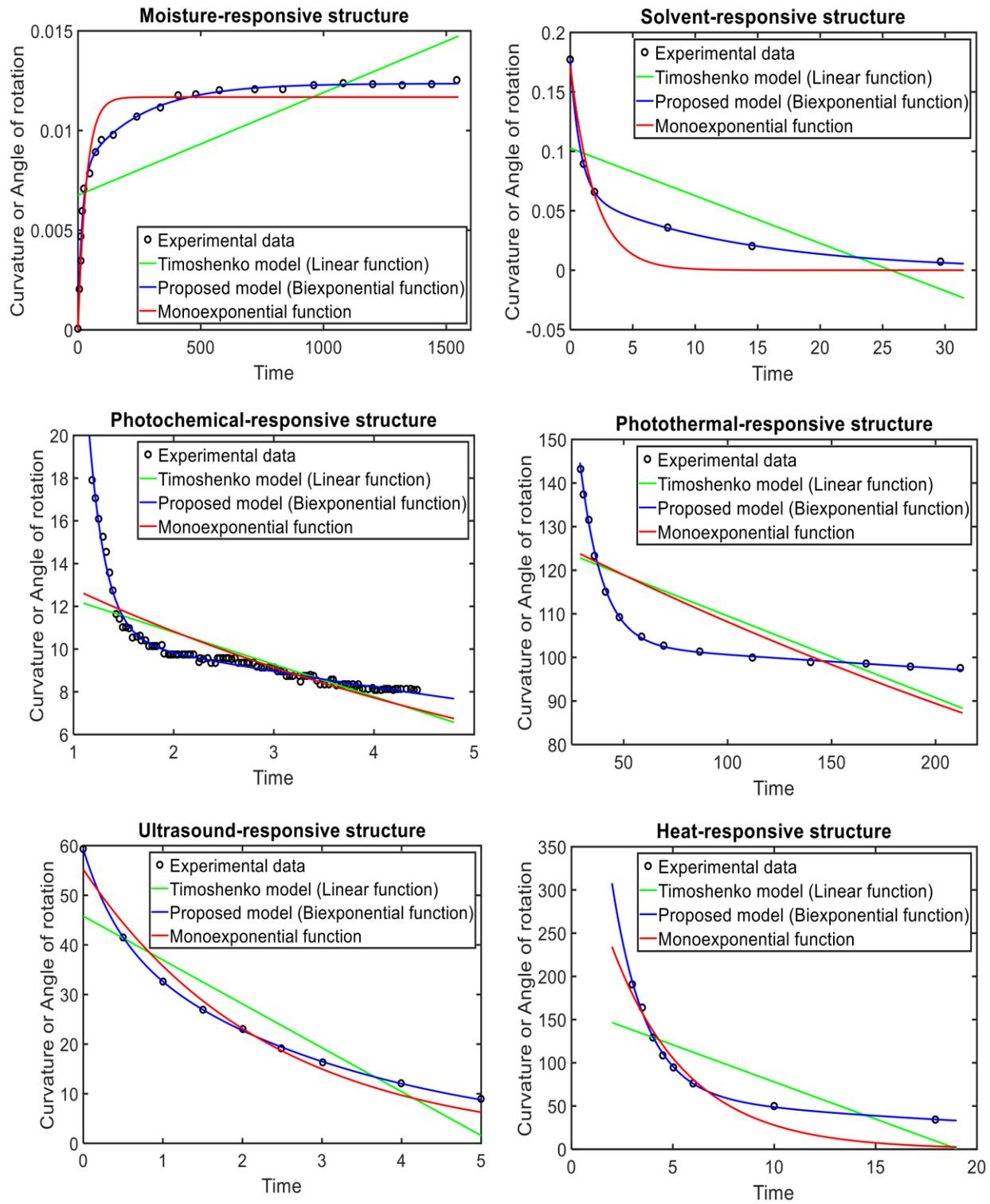

Figure 4. Validation of the proposed model by experimental data from separate studies in the literature [36-41], for both the on and off regions and various stimuli such as moisture [36], solvent [37], photochemical [38], photothermal [39], ultrasound [40], and heat [41]. We performed the curve fitting by Levenberg-Marquardt [42, 43] method. We also tried other least-squares algorithms and obtained similar results. The axes units are eliminated, as the absolute values are not essential to convey the idea.

**Remark 1. True time-dependent behavior of 4D printed multi-material structures.** The results show that generally speaking, two different time-constants govern the time-dependent shape-shifting behaviors of multi-material 4D structures. Nevertheless, in some cases, the two time-constants ($\tau_I$ and $\tau_{II}$) may be approximately equal, and the time-dependent behavior can be modeled by a mono-exponential equation. In some cases, the resistive (viscosity) effect at the interface of the active and passive materials (that is reflected in $\tau_I$) is negligible, and the first exponential term vanishes. In addition, sometimes, the two time-constants are large, and the proposed bi-exponential formula tends to the linear (Timoshenko) model. That is, if $\tau_I$ and $\tau_{II} \to$ large values, then $K_I \left(1 - e^{\frac{-t}{\tau_I}}\right) + K_{II} \left(1 - e^{\frac{-t}{\tau_{II}}}\right) \cong bt$, where $b$ is a constant. This point can be realized by analyzing the related graph or by using Taylor series. Thus, in linear cases, both the proposed bi-exponential model and Timoshenko bimetal model work.

It should be noted that in the heat- and ultrasound-responsive structures of Figure 4, the mono- and bi-exponential models are relatively close to each other.

**Remark 2. Final shape versus instantaneous shape.** We found that the Timoshenko bimetal model [24], as well as the mono-exponential function, cannot capture the correct time-dependent (instantaneous) behavior. However, when the final shape is achieved ($t \to$ large values), both the Timoshenko bimetal model [24] and our formula (equation (1)) would be

$$\kappa_{final} \propto \frac{1}{\frac{h}{2} + \frac{2(E_1 I_1 + E_2 I_2)}{h}\left(\frac{1}{E_1 a_1} + \frac{1}{E_2 a_2}\right)}. \tag{12}$$

Two points should be discussed in this regard. First, the above outcome implies that both the Timoshenko bimetal model and our formula provide similar analyses for time-*independent* behaviors (such as the effect of thickness on curvature). Second, in the literature, some experimental studies reported the decrease of maximum (final) curvature with an increase of layer thickness, while some others reported the increase of maximum (final) curvature with an increase of layer thickness. It is worth digging into this point by an analytical study as the following.

Equation (12) depends on two quantities, Young`s moduli ($E_i$) and layers thicknesses ($a_i$), as $h$ and $I_i$ are functions of layers thicknesses ($h = a_1 + a_2$, $I_i = \frac{1}{12}a_i^3$; based on the basic assumptions of the Timoshenko bimetal model, the width of the strip was assumed to be small and specifically was taken unity as seen in Figure 2). Therefore, we have

$$\kappa_{final} \propto \frac{1}{\frac{a_1 + a_2}{2} + \frac{2}{a_1 + a_2}\left(\frac{E_1}{12}a_1^3 + \frac{E_2}{12}a_2^3\right)\left(\frac{1}{E_1 a_1} + \frac{1}{E_2 a_2}\right)}. \tag{13}$$

By analyzing equation (13), we can find that the curvature first increases and then decreases with increase in $a_1$. Because equation (13) is symmetric in terms of $a_1$ and $a_2$, the curvature has a similar trend with respect to $a_2$. The region, in which the increase or decrease of the curvature happens, depends on relative values of $a_1$ and $a_2$, as well as $E_1$ and $E_2$. Here, we generate some possible scenarios as shown in Figure 5. A similar result to scenario (c) in Figure 5 was proposed by Timoshenko [24]. Due to the symmetry of (13), the same results as Figure 5(a)-(b) are valid if we switch $a_1$ and $a_2$ in these two plots.

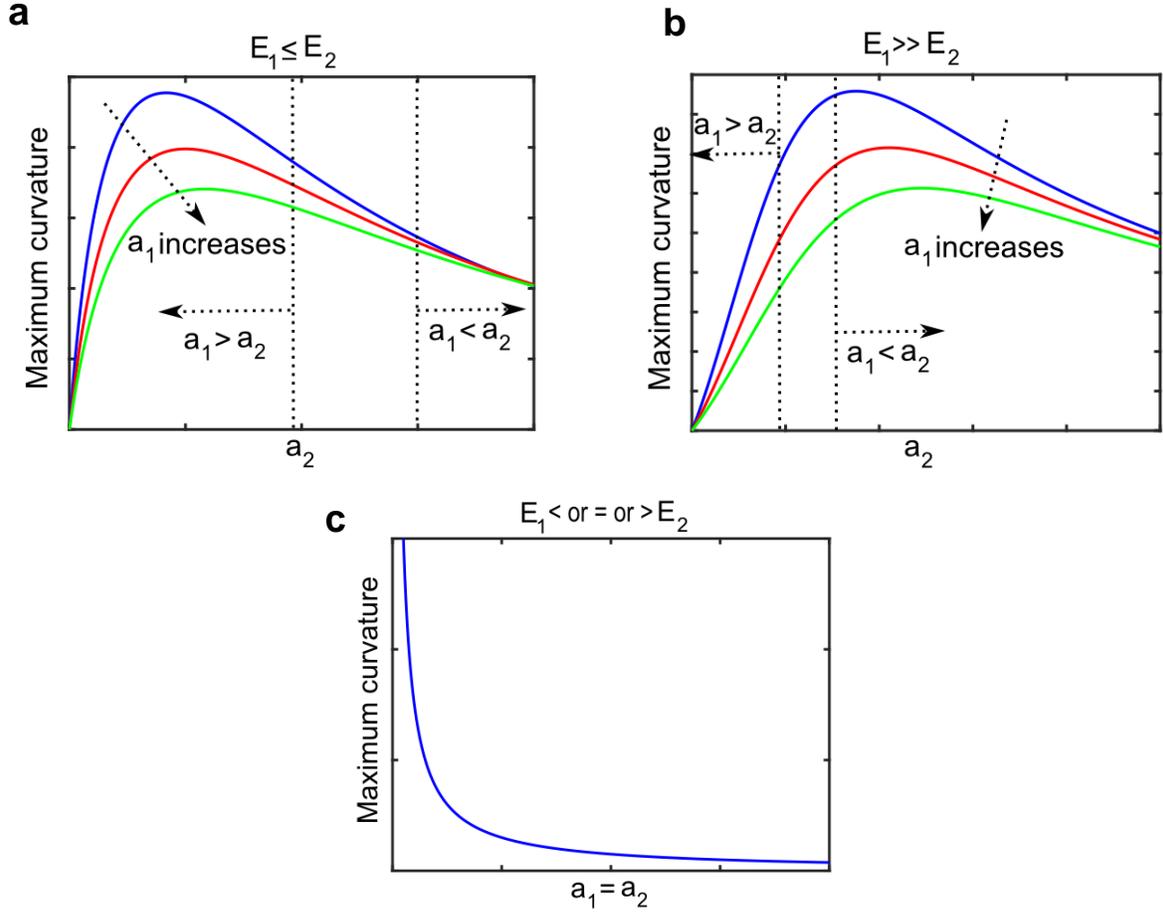

Figure 5. Depending on the relative values of $a_1, a_2, E_1,$ and $E_2$, the relationship between curvature and layers thicknesses would be different (it can be decreasing, increasing, or a mixed behavior). This figure is based on equation (13).

**Remark 3. Shape-shifting speed.** The shape-shifting speed is important almost in any application performing dynamic intelligent behavior over time, and becomes more crucial in some applications such as autonomous deployment in space missions, drug delivery systems, detection devices, and so on. By taking derivatives of equations (1) and (11), the magnitude of the shape-shifting speed for both the on and off regions would be

$$(shape\ shifting\ speed)_{on} = (shape\ shifting\ speed)_{off} =$$

$$\frac{d\kappa(t)}{dt} = \frac{\frac{H_I}{\tau_I}\left(e^{\frac{-t}{\tau_I}}\right) + \frac{H_{II}}{\tau_{II}}\left(e^{\frac{-t}{\tau_{II}}}\right)}{\frac{h}{2} + \frac{2(E_1 I_1 + E_2 I_2)}{h}\left(\frac{1}{E_1 a_1} + \frac{1}{E_2 a_2}\right)} = \frac{K_I}{\tau_I}\left(e^{\frac{-t}{\tau_I}}\right) + \frac{K_{II}}{\tau_{II}}\left(e^{\frac{-t}{\tau_{II}}}\right). \quad (14)$$

As seen in equation (14), the shape-shifting speed is time-dependent with the specific format above. However, the Timoshenko bimetal model [24] gives a constant shape-shifting speed over time.

Based on equation (14), for a large amount of time (i.e., when the final shape is going to be achieved), the shape-shifting speed tends to zero for both the on and off regions. This point can also be captured from Figure 3 as both the on and off regions becomes flat (constant) for $t \to$ large values, and the derivative of a constant function is zero.

**Remark 4. Stimulus power.** Here, we analyze the effect of stimulus power on the time-dependent behavior. By the expression "stimulus power", we mean light intensity, temperature magnitude, pH value, moisture (RH) content, enzyme concentration, current magnitude, solvent concentration, and so on. By analyzing the various parameters of equation (1) (and considering the concepts associated with the second and third laws), we can find that the stimulus power will affect three parameters, $H_I$, $H_{II}$, and $\tau_{II}$. Therefore, the time-dependent behaviors of two different stimulus powers would be

$$\begin{cases} \kappa^{(1)}(t) = \dfrac{H_I^{(1)}\left(1 - e^{\frac{-t}{\tau_I}}\right) + H_{II}^{(1)}\left(1 - e^{\frac{-t}{\tau_{II}^{(1)}}}\right)}{\dfrac{h}{2} + \dfrac{2(E_1 I_1 + E_2 I_2)}{h}\left(\dfrac{1}{E_1 a_1} + \dfrac{1}{E_2 a_2}\right)} = K_I^{(1)}\left(1 - e^{\frac{-t}{\tau_I}}\right) + K_{II}^{(1)}\left(1 - e^{\frac{-t}{\tau_{II}^{(1)}}}\right) \\ \\ \kappa^{(2)}(t) = \dfrac{H_I^{(2)}\left(1 - e^{\frac{-t}{\tau_I}}\right) + H_{II}^{(2)}\left(1 - e^{\frac{-t}{\tau_{II}^{(2)}}}\right)}{\dfrac{h}{2} + \dfrac{2(E_1 I_1 + E_2 I_2)}{h}\left(\dfrac{1}{E_1 a_1} + \dfrac{1}{E_2 a_2}\right)} = K_I^{(2)}\left(1 - e^{\frac{-t}{\tau_I}}\right) + K_{II}^{(2)}\left(1 - e^{\frac{-t}{\tau_{II}^{(2)}}}\right) \end{cases} \quad (15)$$

To illustrate the effect of stimulus power on time-dependent behaviors, we consider five different stimulus powers. The related formulas would be

$$\begin{cases} \kappa^{(1)}(t) = K_I^{(1)}\left(1 - e^{\frac{-t}{\tau_I}}\right) + K_{II}^{(1)}\left(1 - e^{\frac{-t}{\tau_{II}^{(1)}}}\right) \\ \kappa^{(2)}(t) = K_I^{(2)}\left(1 - e^{\frac{-t}{\tau_I}}\right) + K_{II}^{(2)}\left(1 - e^{\frac{-t}{\tau_{II}^{(2)}}}\right) \\ \kappa^{(3)}(t) = K_I^{(3)}\left(1 - e^{\frac{-t}{\tau_I}}\right) + K_{II}^{(3)}\left(1 - e^{\frac{-t}{\tau_{II}^{(3)}}}\right), \\ \kappa^{(4)}(t) = K_I^{(4)}\left(1 - e^{\frac{-t}{\tau_I}}\right) + K_{II}^{(4)}\left(1 - e^{\frac{-t}{\tau_{II}^{(4)}}}\right) \\ \kappa^{(5)}(t) = K_I^{(5)}\left(1 - e^{\frac{-t}{\tau_I}}\right) + K_{II}^{(5)}\left(1 - e^{\frac{-t}{\tau_{II}^{(5)}}}\right) \end{cases} \quad (16)$$

and the general graph would be similar to Figure 6(a). The general trend observed in Figure 6(a) is consistent with the experimental data found in the literature [44, 45, 39, 38].

There might be some applications, in which faster response without any change in the final shape (unlike Figure 6(a)) is desirable. To this end, smaller time-constant ($\tau$) with the same coefficient ($K$) is required. For five different scenarios, the related formulas would be

$$\begin{cases} \kappa^{(1)}(t) = K_I\left(1 - e^{\frac{-t}{\tau_I}}\right) + K_{II}\left(1 - e^{\frac{-t}{\tau_{II}^{(1)}}}\right) \\ \kappa^{(2)}(t) = K_I\left(1 - e^{\frac{-t}{\tau_I}}\right) + K_{II}\left(1 - e^{\frac{-t}{\tau_{II}^{(2)}}}\right) \\ \kappa^{(3)}(t) = K_I\left(1 - e^{\frac{-t}{\tau_I}}\right) + K_{II}\left(1 - e^{\frac{-t}{\tau_{II}^{(3)}}}\right), \\ \kappa^{(4)}(t) = K_I\left(1 - e^{\frac{-t}{\tau_I}}\right) + K_{II}\left(1 - e^{\frac{-t}{\tau_{II}^{(4)}}}\right) \\ \kappa^{(5)}(t) = K_I\left(1 - e^{\frac{-t}{\tau_I}}\right) + K_{II}\left(1 - e^{\frac{-t}{\tau_{II}^{(5)}}}\right) \end{cases} \quad (17)$$

and the general graph would be similar to Figure 6(b). In order to tune the response speed without interfering with the final shape, separate studies are needed. Carbon nanotubes (CNTs) [46] may be a possible solution, as they can be incorporated into stimuli-responsive materials to tune their response speed [47].

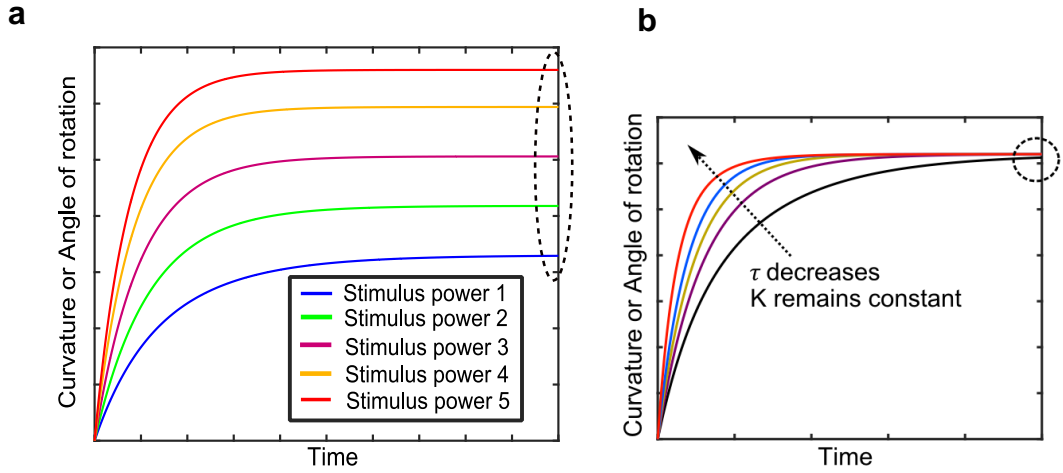

Figure 6. (a) The general effect of stimulus power (e.g., light intensity, pH value, temperature magnitude, and so on) on time-dependent behavior. This plot is based on equation (16). (b) Tuning the response speed, without changing the final shape. This plot is based on equation (17).

**Remark 5. Extension of the developed concepts to complicated multi-material 4D structures.** Based on the concepts developed here, one can analyze, predict, and tune the time-dependent behaviors of multi-material 4D structures almost at any level of complexity. Let us consider one example exhibited in Figure 7. This case has four different types of materials, two of which are active materials. For the top two layers, we have one exponential term of type I and one exponential term of type II. Similarly, we can consider the middle two layers, and the bottom two layers. By superposing these terms, this multi-material structure has three exponential terms of type I and three exponential terms of type II (nevertheless, some of these exponential terms can be equal in a specific case, as discussed in Remark 1). In addition, this case will have the same general graph shown in Figure 3, as its governing equation is a summation of exponential terms. However, the slope of its graph would be different at various points (i.e., it will have steeper slopes in some regions).

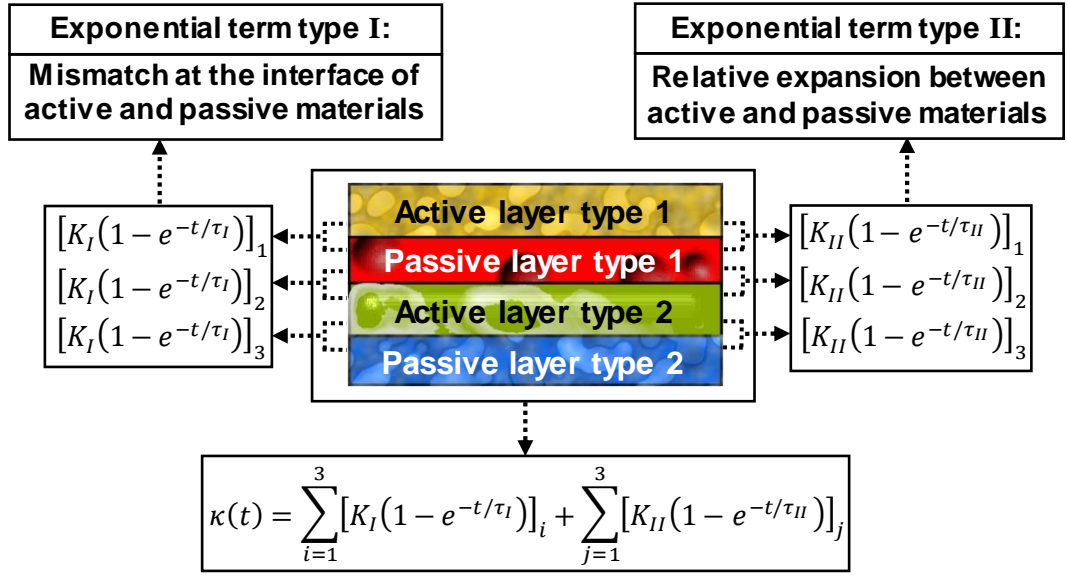

Figure 7. A 4D structure with more than two types of materials.

To realize the complicated 4D structures, in addition to the multi-material (rather than two-material) structures discussed above, two more important points should be taken into consideration. First, our model provides the time-dependent behavior of the "curvature" that is a fundamental building block of shape-shifting in multi-material structures. Other higher-level shape-shifting quantities such as curling, twisting, coiling, and their combinations originate from this quantity. Second, we have discussed the time-dependent behavior of the curvature in one direction (narrow strip). For a plate having the same materials as the original narrow strip, the curvature would be the same in any direction (this point can be concluded by analyzing [24]), and can be modeled by the same two exponential terms of the original (parent) narrow strip. For a plate that has different materials in different directions, the curvature would be different in each direction. However, the same two types of time-constants and exponential terms proposed in this study can be used to model the time-dependent curvature in a specific direction, accordingly. Future studies may incorporate the proposed two time-constants in extensions to plates and so on.

It should also be noted that the concepts have been developed for multi-*material* structures that may not necessarily be multi-*layer*. In some cases, the boundary between active and passive layers may not be as clear as Figure 2. However, in these cases, eventually, the active and passive materials will have contact in some regions, and the same concepts developed here will be present.

**Remark 6. Shape memory effect (SME).** As we touched on earlier, SME is *not* an intrinsic property [9]. The readers are referred to ref. [9] for more information on the SME and shape memory polymers. The four shape-shifting mechanisms discussed and quantified in the second law can take place with or without SME. The key point is that these four shape-shifting mechanisms are the underlying physical concepts for relative expansion and subsequent shape-shifting in "multi-material" 4D structures. For example, the "heat-responsive structure" of Figure 4 illustrates a typical shape memory polymer in a bilayer with a passive material, and its shape-shifting in the "multi-material" structure is enabled by the relative "thermal expansion" of the active and passive materials as mentioned by Zhang et al. [41, 48].

Because of the aforementioned discussion, we did not put SME as one category of fundamental physical concepts in the second law. Nevertheless, the following discussion can be useful.

Most of the shape memory polymers are thermo-responsive, and to enable their shape memory effect, their temperature should go beyond a critical temperature, e.g., glass transition temperature (whether by direct heat or other stimuli such as light, electricity, and electromagnetic field to change the temperature, indirectly) [9]. It is worth mentioning that the researchers [48, 49] reported that the "strain-time" relationship of a shape memory polymer above its glass transition temperature could be modeled by the exponential formula $\varepsilon(t) = C\left[1 - \exp\left(-\frac{t}{\tau}\right)\right]$.

**Remark 7. Other manufacturing processes.** We distilled three laws that govern the shape-shifting behaviors of almost all the 4D multi-materials, whether fabricated by additive manufacturing (AM) and so-called 4D printed structures or made by other manufacturing processes.

Stimuli-responsive multi-materials can be made by various manufacturing processes; however, AM has some benefits. Two general advantages are discussed here. First, the same reasons that motivate us to use AM for conventional (passive) materials, will be motive for utilizing AM for stimuli-responsive (active) materials. In other words, 4D printing conserves the general advantages of AM (such as material waste reduction, elimination of molds, dies, and machining [50], and providing complex geometries) that are not present in other manufacturing processes. In addition, unlike other manufacturing processes, 4D printing provides an encoded multi-material smart structure in a single run. Second, AM helps us to manipulate the "structure" of multi-materials, precisely, to enable various shape-shifting behaviors. In other words, 4D "printing" enables encoding local anisotropy [5] in multi-materials.

In fact, before the initiation of the 4D printing idea, researchers were *not* usually trying to find a specific printing path by mathematics that could yield a predictable and desired shape-shifting over time. 4D printing is a new manufacturing paradigm that combines smart materials, mathematics, and multi-material additive manufacturing, as Momeni et al. [6] organized it in a systematic way (Figure 8).

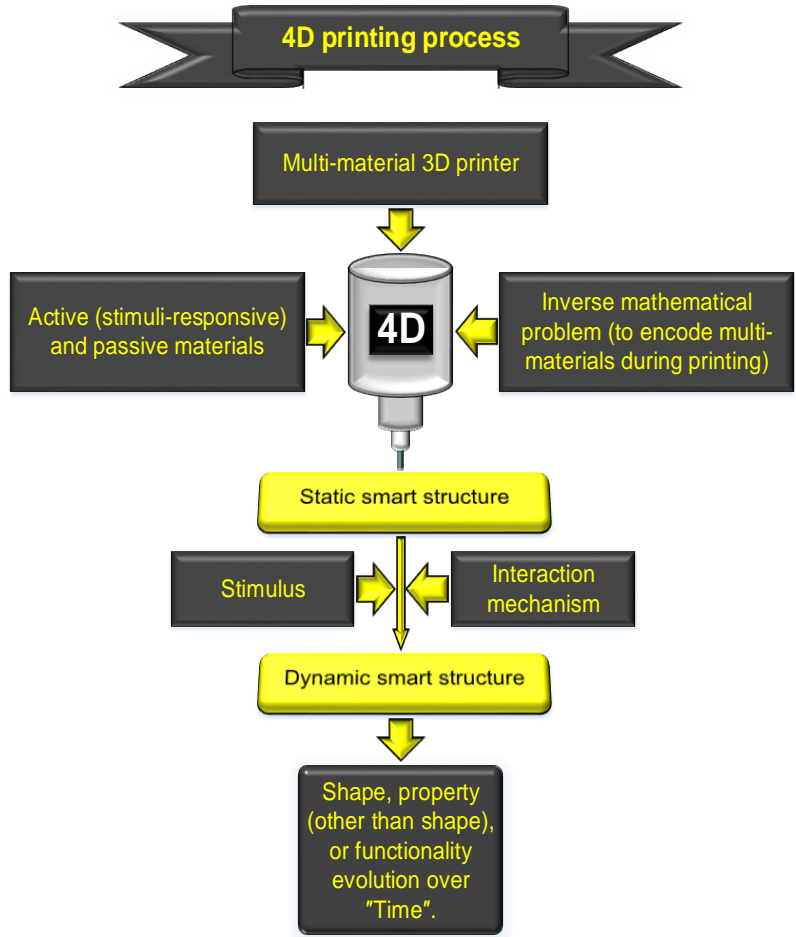

Figure 8. 4D printing process (Momeni et al. [6]).

**Remark 8. Scope and exceptions.** Throughout this work, we did not make any specific assumption regarding the types of materials, stimuli, and length scales, for which these three laws are valid.

However, three points should be mentioned about the scope of this study. First, these three laws aimed at "shape-shifting" behaviors that are currently the focus of studies in the 4D printing field. The evolutions of other properties such as color or thermal resistance have not been discussed in this study, although they are of interest. Nevertheless, the shape-shifting behavior can provide evolutions in other properties or functionality. As an example, the researchers [51, 52] demonstrated smart patterned surfaces that can alter their geometry in a manner that leads to changes in their effective emissivity to eventually control the satellite temperature, without using controllers and energy supplies. Second, these laws are associated with "multi-material" structures having active and passive materials. Third, simple linear expansion/contraction shape-shifting behaviors are excluded from this study.

In science and engineering, laws are flexible and can have exceptions. We have inspected more than 200 related published works, whether stimuli-responsive structures fabricated by printing or made by other manufacturing processes. However, we could not find any exception (counterexample) for our laws (considering the remarks and scope of this study). Nevertheless, we put the expression "almost all" in our three laws for possible exceptions in the future. By the way, our results are general and also important, as they target the 4th D of 4D multi-materials. Some of the exiting works, at the first glance, may seem counterexample for these three laws. However, by in-depth analysis and considering their fundamental physics, their compliance with these three laws will be comprehended. For example, the built-in (direct) 4D printing proposed by some researchers [53, 54] has the same "underlying physics" of conventional shape memory polymers. As we touched on earlier, to enable the SME of SMPs a mechanical force is required in addition to heat (the thermomechanical (thermo+mechanics) cycles of SMPs indicate this point, as well). In the built-in (direct) shape memory effect [53, 54], this mechanical force (that is used for programming) is provided during the printing, i.e., the printing and programming steps are integrated. A similar concept has been discussed in the introduction and has also been mentioned by some other studies [22, 41]. The underlying physics of shape-shifting of these examples in multi-material structures is the same as "heat-responsive structure" of Figure 4 that has been discussed in Remark 6.

Sometimes, a formula is derived and validated; however, it is valid and applicable only for a specific range of cases. The bi-exponential formula derived and validated here is a universal governing equation that can model and predict the 4th D of any 4D multi-material structure as seen in Figure 4, this is because we built its bases, comprehensively. We use the word "law", because the related results are general and are also required for understanding, modeling, and predicting the shape-shifting behaviors of multi-material 4D printed structures.

**Remark 9. Future directions and opportunities.** A market study announced on the Reuters website [55] indicates that 3D printing market will have a CAGR (compound annual growth rate) of 30.20% from 2017 to 2022, while 4D printing market will have a CAGR of 40.30% from 2017 to 2022. Definitely, the total investment on 3D printing is more than 4D printing, at the present time. However, the aforementioned statistics compare the growth (the future perspective) between 3D and 4D printing.

3D printing, by itself, is considered a multi-disciplinary field. Thus, more research areas will be involved in the 4D printing field. This diversity can increase the strength of 4D printing.

Future works can consider several topics. One of the main topics is the compatibility of materials in multi-material structures. The materials should form a strong bond at their interface. Their bond should also remain strong under stimuli. The other topic is measurement and modeling of some of the parameters in the bi-exponential formula such as the time-constants. Some of the parameters should be measured for active-passive materials, while models (whether case-specific or general) can be developed for some of them. Experimental studies can be conducted to find the exact values or the ranges of parameters for categories of materials. The next topic lies in software and hardware developments and their "integration". Future 4D printing software should have some levels of predictions. It can also provide a situation for tuning the behavior over time. Future 4D printing hardware developments require some controls strategies that can handle multi-material printing. Assume that ten different materials are going to be encoded in a single-piece structure, in its various locations (voxels), through ten different nozzles that are going to work, simultaneously. Printability of smart materials is the next work. There are many smart materials; however, they need to become printable. Tuning the response speed is also an important topic. The next point is related to product development by 4D printing. 4D printing is not just a concept, it is also a manufacturing paradigm and manufacturing is always connected with design and their integration leads to a product. Therefore, new products or applications that can have unique features by 4D printing, should continuously be considered and addressed.

**Remark 10. Closing.** As a summary of the proposed three laws, Figure 9 is presented.

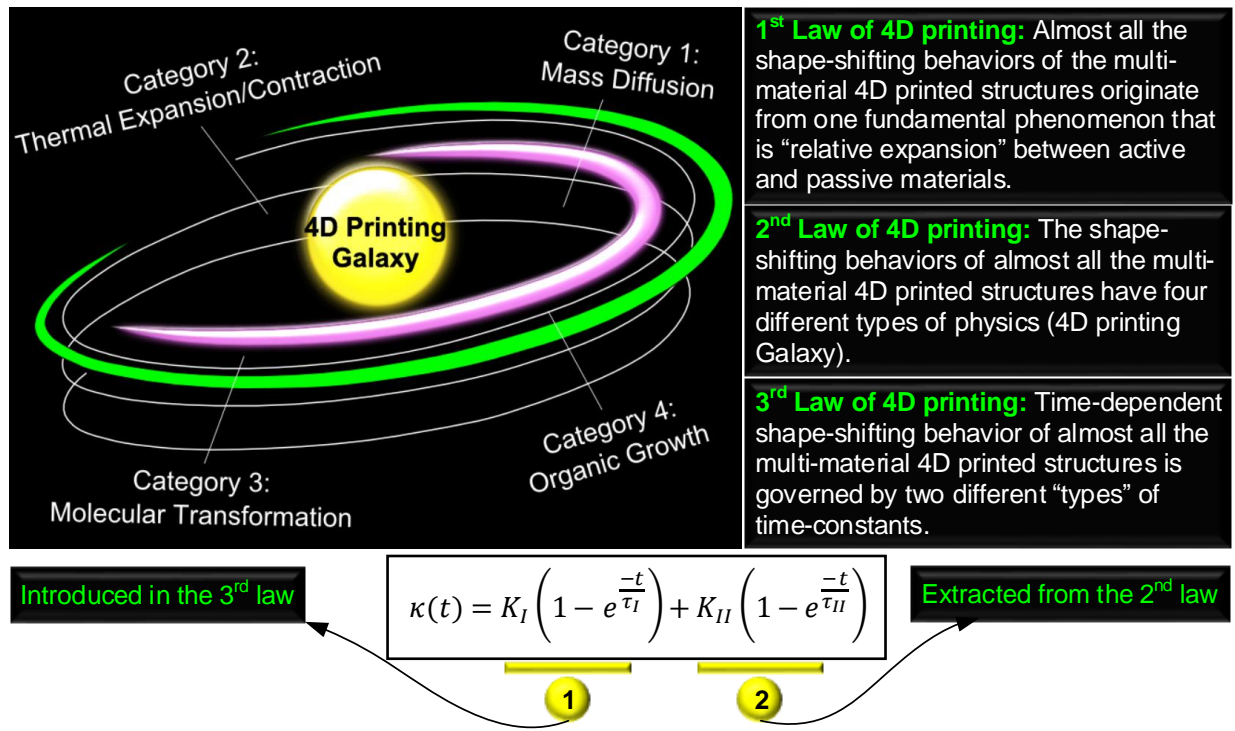

Figure 9. A summary of our 4D printing laws. (the galactic shape of this figure has been inspired by a display designed by Rod Hill, showing advancements in Reconfigurable Manufacturing Systems, and installed on the wall of the ERC-RMS Center at the University of Michigan.)

## 3. Conclusions

Stimuli-responsive materials have demonstrated their promising applications. Any emerging application that is enabled by functional and stimuli-responsive materials can be elaborated in the field of 4D printing due to the unique attributes of the multi-material additive manufacturing process. Here, by a detailed, but systematic, qualitative and quantitative study, we generated and validated a bi-exponential formula that governs the shape-shifting behavior of almost all the multi-material 4D structures over time. We showed that two different types of time-constants *are needed* to capture the correct time-dependent behavior of 4D multi-materials. The purpose of the first and second laws was to understand the 4$^{th}$ D of 4D printed multi-material structures, and the purpose of the second and third laws was to model and predict the 4$^{th}$ D. The results of this study can serve as a guideline and general design principles for the future. They can also be incorporated into future software and hardware developments. We should note that pure experimental study might not be able to generate a general conclusion for the relationship between two quantities, as the experimental study may not cover all the possible regions of the relationship in various cases. Drawing a systematic conclusion is the strength of analytical study (validated by experimental data) as done here. Recently, many exciting works have been demonstrated by 4D printing that can hardly be achieved by other processes. However, more collaborations between scientists and engineers from various fields are needed to move from lab to fab and unveil full potentials of 4D printing. In history, some topics are coined and enter into research communities and become popular, then after a short period, they lose the broad interests. However, it is expected that 4D printing remains attractive and useful for a long period because stimuli-responsive materials have already demonstrated their promising applications in various fields. Furthermore, 4D printing helps us to locally and precisely encode the stimuli-responsive multi-materials by leveraging the strengths of multi-material additive manufacturing and mathematics that are the elements of 4D printing. The general equation that we obtained here is the starting point and can be further developed and incorporated into future 4D printers.

**Author contributions**

F.M. conceived the study, obtained & analyzed the results, and wrote the manuscript. J.N. discussed the results, reviewed the manuscript, and provided feedback.

**Acknowledgments**

F.M. wants to thank S.M. Wu Manufacturing Research Center in the Mechanical Engineering Department at the University of Michigan-Ann Arbor.